\documentclass[12pt,preprint]{aastex}

\def\etal    {et al.}
\def\oceti   {$o$~Ceti}
\def\chicyg  {$\chi$~Cygni}
\def\gmpercc {gm~cm$^{-3}$}
\def\Msun    {M$_\odot$}

\slugcomment{Submitted to Ap.J.;revised October 19, 2001}

\shorttitle{Mira Light Curves}
\shortauthors{Reid \& Goldston}

\begin{document}


\title{How Mira Variables Change Visual Light 
       by a Thousand-fold}


\author{M. J. Reid and J. E. Goldston}
\affil{Harvard--Smithsonian Center for Astrophysics,
    60 Garden Street, Cambridge, MA 02138}
\email{reid@cfa.harvard.edu, jgoldston@cfa.harvard.edu}



\begin{abstract}

Mira variables change visual light by up to 8 magnitudes over their 
roughly yearly cycle.  Here we present a simple explanation for the 
extremely large amplitudes of light curves of oxygen-rich Mira variables. 
Metallic oxides, such as TiO, form throughout the stellar atmosphere 
as the star cools when approaching minimum light.   When this happens,
the visual light can be almost completely absorbed at large radii, 
extending the visual photosphere to nearly twice its nominal size.  
At these large radii, temperatures can fall to $\approx1400$~K and 
essentially all of the star's radiation emerges in the infrared.  
Since almost no optical light is emitted at these low temperatures, 
Mira variables can decrease their visual light by more than a 
thousand-fold and almost ``disappear'' to the human eye.

\end{abstract}



\section{Introduction}

	Mira variables have long fascinated both amateur and professional
astronomers.  In the late 1500s, European and Asian astronomers noted the 
``disappearance" of a star, most likely \oceti, over a period of months.  
While this star dimmed dramatically, it did not really disappear and, in 
the early 1600s, Fabricus and Holwarda recognized that \oceti\ was a 
variable star \citep{hof97}.  Because of the extreme changes of its 
brightness, \oceti\ was called Mira (or ``marvelous one'').  Indeed, Mira 
variables are marvelous; some regularly change visual luminosity by
$\approx8$ magnitudes, or factors of more than 1000, over periods of hundreds 
of days \citep{mer40}.  

	Mira variables are evolved red giants with extraordinarily
large sizes; a typical Mira has a radius of $\approx2$~AU 
(e.g., Pease 1931; Labeyrie \etal\ 1977; Lopez \etal\ 1997; Tej \etal\ 1999). 
Mira variables are thought to pulsate, owing to opacity from the variable 
ionization of hydrogen and helium below the stellar surface 
(e.g., Keeley 1970).
These pulsations emerge as shock waves and propagate upwards into 
the lower photosphere near maximum visible light (e.g., Slutz 1976).
A Mira variable can have a very
complex, dynamic atmosphere, and much theoretical and computational 
work has been expended attempting to understand the physical processes
involved (e.g., Wood 1979; Bertschinger \& Chavalier 1985; Bowen 1988;
Bessel \etal\ 1989; Plez, Brett \& Nordlund 1992; H\"ofner \etal\ 1998;
Woitke \etal\ 1999; Helling, Winters \& Sedlmayr 2000).
Because Miras are so large, their surface temperatures 
are low and many molecules form in their atmospheres.
At distances of a few stellar radii, temperatures fall even lower, which
allows the formation of silicate or graphite ``dust."
Once dust grains form, radiation pressure rapidly disperses them into 
interstellar space, enriching the environment with heavy elements.  
The ability of red giant stars to form and disperse circumstellar
material is crucial to the life-cycle of stars and for life as we know it.  
Thus, the study of the workings of the cool atmospheres of such stars has 
great significance to astrophysics. 

     Nearly 70 year ago, \citet{pn33} showed
that Mira variables have temperature variations of $\approx30$\%
and total luminosity variations of factors of $\approx2$ over their roughly 
yearly cycle.  The luminosity, $L$, the effective temperature, $T_{eff}$, 
and the radius, $R$, of a star are related by 
$L = \sigma T_{eff}^4 \pi R^2$,
where $\sigma$ is the Stephan-Boltzmann constant.      
Using this relation, measurements of $L$ and $T_{eff}$ give an 
estimate of $R$, and radii of Miras determined in this manner indicate 
size changes of $\approx40$\%.  

The observed visual \citep{mat01} and infrared light \citep{lw71} 
curves of \chicyg, a bright oxygen--rich Mira variable, 
are presented in Fig.~1 as {\it green} and {\it red crosses}, 
respectively.  Plotted with solid lines are the expected flux 
density as a function of time for a time--variable black-body model
based on changes in temperature and size as determined by \citet{pn33}.
The infrared model falls only about 1 magnitude dimmer than the data 
and has approximately the correct amplitude.  Thus, by adjusting the 
stellar size, this model could fit the infrared light curve well.  However,
the visible curve ({\it green}) starts at maximum light about 2 magnitudes 
too bright, and the amplitude is about 4 magnitudes too small.
This is a general problem for Mira variables:
the measured temperature and radius changes, determined mostly from 
infrared data, predict only about 3 to 4 magnitudes of visual light variation, 
whereas up to 8 magnitudes are sometimes observed.

Good fits to the observed light curves cannot be obtained by simply
changing the variation of the temperature and radius of a stellar 
black-body.  If the black-body temperatures are allowed to decrease
considerably below those indicated by infrared data, one can
generate a very large visual amplitude and reproduce the visual
magnitude at minimum light.  For example, in Fig.~1 we show 
light curves ({\it dashed lines}) for a stellar black-body
whose temperature varies from 2300~K at maximum light to 1300~K
at minimum light.  While the visual light curve can be successfully
modeled in this manner, the model for the infrared light curve degrades
considerably.  The problem is that time varying black-body models 
cannot simultaneously fit the optical and infrared light curves of a 
star like \chicyg.  

Perhaps one could explain Mira light curves through the effects of
a strong opacity source that varies considerably with wave-band.  
Astronomers have long suspected that molecular absorption in the 
atmospheres of Mira variables contributed to the large change in visual 
luminosity (e.g., Pettit \& Nicholson 1933; Smak 1964; Celis 1978).
However, to our knowledge, this problem has never been directly addressed.
In this paper, we present an explanation of how the formation of
metallic oxides (notably TiO), as the star periodically cools,
can solve the problem of Mira light curves.

\section{Basic Physical Explanation}
 
     The fact that Mira variables change infrared (and total) luminosity
by only about 1 magnitude suggests that the star is not undergoing
exotic changes.  Where then should one look to explain the dramatic
optical variations?  Consider a star with an average
temperature of 2200 K and a radius of 2 AU.  At a distance of 100 parsecs, 
this star would have visual magnitude, $m_v$, of $+4.5$~mag.  
Varying the radius of such a star by 
$\pm20$\% (from 2.4 to 1.6 AU) changes $m_v$ by only $\pm0.4$~mag.  
Clearly, size changes are
insufficient to explain the large observed swings in visual light.  
On the other hand, varying the temperature of the star by only $\pm15$\% 
(from 2500 K to 1900 K) strongly affects visible light --- changing 
it by 3.4 mag.   While this is 
inadequate to explain the large changes in the visual light of many Miras,
it does suggest that the explanation comes predominantly 
through the effects of temperature variations.

     The emission from a Mira variable with an 8 magnitude amplitude could be 
explained if, at minimum light, the radiation comes from material at about 1400 K.
How far above the stellar surface\footnote{Throughout this paper we use the 
terms stellar surface, stellar size, and stellar radius to mean those 
quantities measured at infrared wavelengths in line-free regions.} 
does this temperature occur?   
If, for example, the temperature decreases with radius as $r^{-1/2}$, then 
a star with a minimum surface temperature of 1900 K 
will experience temperatures near 1400 K at about 1.8 stellar radii ($R_*$).
Thus, the problem of explaining the large amplitude of Mira variables may be 
reduced to finding a source of visual opacity sufficient to make the star 
appear 80\% larger than the stellar size at minimum light, while remaining 
close the stellar size at maximum light.  In order to accomplish this, the
opacity in the stellar atmosphere should have the following 
characteristics: 1) The opacity must increase significantly as the
temperature decreases from about 1900 to 1400 K.  
Were, for example, the opacity to remain high throughout
the stellar cycle, the amplitude of the variations would only approach
about 5 magnitudes.
2) The visual opacity must be great enough to achieve optical depths of $\sim1$
for the low densities expected in the stellar atmosphere at $r \approx 2R_*$.  

     It has long been known that oxygen-rich (M-- and S--type) Miras have 
strong metallic oxide absorption bands across the visible spectrum, with
titanium oxide (TiO) typically providing the dominant absorption \citep{mer40}.
In the extended photosphere of an oxygen rich Mira variable, 
the chemical reaction of Ti + O $\rightleftharpoons$ TiO  
proceeds from predominantly atomic to molecular phases for temperatures 
between about 2000 and 1600~K for total densities of $10^{-10}$
and $10^{-14}$~\gmpercc, respectively.
So as the star cools when approaching minimum light, titanium can
transition from mostly atomic to molecular form
in the extended atmosphere of a Mira variable.  
Once TiO forms, it is extremely efficient at absorbing optical photons.  
This is why some sun screen products use titanium (or other 
metallic) oxides.  

     The implications for a star that naturally forms sun screen in its 
atmosphere can be dramatic.  The optical depth to molecular absorption 
can be written as follows:  $\tau = \int n_m~\alpha~d\ell~~,$
where $n_m$ is the molecular density, $\alpha$ is the absorption 
cross-section, and $d\ell$ is path length along our line-of-sight.
Can one achieve $\tau\sim1$ in the extended photosphere owing to 
TiO absorption?  For an approximate estimate of opacity, assume a 
characteristic path length of $\ell \sim 10^{13}$~cm and a TiO cross-section, 
averaged over the visual band, of 
$\alpha\sim 10^{-17}~{\rm cm}^2$ \citep{jor94}.  
Then unit optical depth requires a characteristic TiO number density of
$n_m = 10^4~{\rm cm}^{-3}$.  For solar abundances, this corresponds
to a total mass density of $\sim10^{-13}$~\gmpercc.
Such a mass density is reasonable in the extended 
stellar photosphere of a Mira variable.  Thus, near minimum light
TiO (and other metallic oxides) can absorb most of the optical light 
and produce a ``visual surface" far above the true stellar surface.
  
The change in the visual appearance of a Mira between maximum and
minimum visible light is schematically represented in Fig.~2.  
The left-hand panel of the figure represents the star near maximum
visual light ($\phi=0$); one can see through the extended atmosphere, in
which metals such as Ti are predominantly in atomic form, almost down to the
stellar surface. (Actually, even though the fraction of Ti in molecular form
is very low at temperatures above 2000 K, a reasonable amount of TiO 
is found near the stellar surface, owing to the extremely high densities 
encountered there.)
The right-hand panel of Fig.~2 represents the star near minimum light 
($\phi\approx0.5$); here the temperature of the star has declined and TiO 
is abundant in the extended photosphere.  
At this point, one can only ``see'' down to a radius of $\approx1.8R_*$.   
When this happens the visual intensity drops dramatically, which can 
explain why Mira variables change visual light by factors exceeding 1000.

\section{Model Light Curves}

     It is exceedingly difficult, and probably beyond current
capabilities, to model precisely all of the physical and chemical
processes in the atmosphere of a Mira variable.  However,
in order to validate the basic physical explanation for Mira
light curves, we need only approximate or representative 
light curves.  This can be accomplished
provided the temperature and density as a function radius and 
optical phase is available.  We found four papers in the recent 
literature that contain time-dependent models of oxygen-rich Mira 
variables and for which temperature and density profiles are 
plotted for maximum and minimum visual light (Bertschinger \& Chavalier
1985, Fig.~13; Bowen 1988, Fig.~6; Bessel \etal\ 1989, Figs.~2 \& 4; 
H\"ofner \etal\ 1998, Fig.~9).
These profiles were digitized and used as input physical models for 
calculations of apparent visual magnitudes.
We note that there are very large differences in
the temperature and density profiles among these models. 

    We used the following procedure to calculate apparent visual magnitudes.
At each radius in the extended photosphere, we evaluated the chemical 
reaction for forming molecular TiO, assuming solar abundance ratios 
([Ti]/[H] = $9.8\times10^{-8}$ and [O]/[H] = $8.5\times10^{-4}$), 
a TiO binding energy of $-6.9$~eV, and chemical equilibrium.  
Since most of the oxygen is probably bound in CO and other molecules, 
we assumed that only about 1\% of oxygen is available to bond to titanium.  
The absorption cross-section of TiO as a function of wavelength was estimated, 
based on calculations provided by R. Kurucz, for a range of 
temperatures and pressures appropriate for the conditions in the outer 
atmosphere of a Mira variable.  The dependence of the cross-sections 
on pressure is insignificant, but the dependence on 
temperature is noticeable, and we interpolated the cross-sections 
in temperature between values calculated at 2200 and 1400 K.

    We calculated numerically the transfer of radiation through the 
atmosphere using a method similar to that described by \citet{rm97}.  
At a given wavelength, we integrated the intensity contribution at each 
point along our line-of-sight toward the star, diminished by absorption 
along the path to the observer.
This integration was repeated for different stellar impact 
parameters, and a monochromatic flux density was determined by summing
the integrated intensity times the solid angle of an annulus appropriate
for each impact parameter.  Next we repeated the integrations over a range of 
wavelengths to sample the entire optical spectrum.  Finally, the total
flux density was obtained by summing over all wavelengths, accounting for 
the transmission characteristics of a visual filter \citep{joh52}.  
In this manner, we generated an apparent visual magnitude, $m_v$, 
for a given stellar phase.  

      The calculated apparent visual magnitudes at maximum light,
using the published temperature and density profiles from the four
papers cited above, spanned a range of about 10 magnitudes.
The model of \citet{bow88} returned $m_v = 6$, which
closely matched the \chicyg\ data at maximum light 
(assuming a distance of 136 pc; see discussion in next section).
Other models typically yielded visual magnitudes a few magnitudes brighter.
At minimum light, the model of Bowen resulted in $m_v = 9$,
about 4.5 magnitudes too bright.  
The differences in visual light between maximum and minimum for most of the
models were only about 3 magnitudes, and for two of the four models the light 
at minimum {\it exceeded} that at maximum.  Thus, these published temperature 
and density profiles cannot explain the visual light curve of
a large amplitude Mira variable like \chicyg.
This should not, perhaps, come as a surprise.  The visual light
is only a small fraction of the total light from the star, and
thus it is relatively unimportant in determining the physical and dynamical 
conditions in the atmospheric models.  
Since the published models have {\it not} been optimized to fit
visual light, it is not surprising that current models fail in this
regard.  However, the visual light is a sensitive probe of
physical conditions in the atmosphere of a Mira variable, and we 
hope that future models would be optimized to fit visual light curves.

     Since published models seem inadequate for our purposes, 
we constructed a simple hydrostatic model for the stellar atmosphere 
and evaluated the transfer of radiation numerically as described above.
This model is computationally manageable and its parameters are easily 
adjusted to attempt to fit light curves.  The model assumes a star 
whose surface temperature and radius vary out of phase as follows:
\begin{mathletters}
\begin{eqnarray}
   T_* = T_0 + \Delta T \cos( \phi )~~~, \\
   R_* = R_0 + \Delta R \cos( \phi + \pi )~~~,
\end{eqnarray}
\end{mathletters}
where $\phi$ is the visual stellar phase.  
The phase shift of $\pi$ in Eq.~(1b), forces the radius to be a minimum
when the temperature is maximum \citep{pn33}.
This star is ``embedded'' in an extended photosphere. 
Following \citet{rm97}, we assume that temperature as a function of radius, 
$T(r)$, in the extended photosphere decreases as follows:
\begin{equation}
   T(r)^4 = T_*^4~\bigl(1 - \sqrt{1-(R_*/r)^2}\bigr)~.  
\end{equation}
This temperature profile decreases approximately as
$r^{-1/2}$ (for $r\ge1.2R_*$) and, for hydrostatic equilibrium, results in a 
density profile, $n(r)$, decreasing with radius given by 
\begin{equation}
   n(r) \approx n_* e^{-C(1-(R_*/r)^{1/2})}~. 
\end{equation}
In Eq.~(3),  $C=2GM_{eff}m_H/kT_*R_*$,
$M_{eff}$ is the effective mass of the star, $m_H$ is the hydrogen mass,
$k$ is Boltzmann's constant, and $G$ is the gravitational constant.
Examples of density and temperature profiles (for parameters given in \S 4) 
are shown in Fig.~3.

\section   {Observed and Modeled Light Curves for \chicyg}

	We chose to model the star \chicyg\ because 
1) it displays a very large, consistent, change in visual light, and
2) excellent visual \citep{mat01} and infrared \citep{lw71} 
light curves are available.  We adopted a distance to \chicyg\ of 136 pc, based
on the period--luminosity relation of \citet{fea89} and the photometry
of \citet{hst95}, and an effective stellar mass of $1~M_\odot$.
We modeled the star as described in \S3, adjusting the parameters of the model 
to fit the optical and infrared light curves.  
A good fit to the light curves of \chicyg\ was obtained for a mean temperature and 
radius of $T_0=2200$~K and $R_0=2.1$~AU, respectively, for the central star.
Changes in temperature, $\Delta T$, and radius, $\Delta R$, over the stellar 
cycle were set to $\pm20\%,$ and a visual extinction of $0.6$~mag was used.
The total mass in the extended atmosphere was a held constant over the
stellar cycle by adjusting the density parameter $n_*$.

     We present the results of the radiative transfer calculations in Fig.~4.
In addition to the visual light curve, we also required the model to 
reproduce the infrared light curve at a wavelength of 1.04 $\mu$m.  
At this wavelength, molecular opacity is very low \citep{lw71}, 
and the infrared radiation comes from the stellar surface.  
As one can see, this model successfully reproduces both the small 
variation in the infrared (where TiO opacity is low) and the extremely 
large changes in visual light.  This demonstrates that, for reasonable 
physical parameters, one can explain the ``marvelous" behavior of a
Mira variable as outlined in \S 2.

     \citet{rm97} measured the radio continuum emission from a sample of 
Mira variables and found that the radio emission comes from $\approx2R_*$.
This is the same region of the extended photosphere that is optically thick 
at visual wavelengths, owing to molecular opacity, near minimum light.  
Since the radio opacity is from an entirely different source 
(H$^-$ free--free interactions) than the optical opacity, one can check 
how well our simple model predicts photospheric emission at radio wavelengths.
Thus, we followed the procedures described by Reid \& Menten and calculated 
the emission at a wavelength of 3.6~cm that would result from the model 
for the optical and infrared emission described above.
The predicted radio emission from the model of \chicyg\ 
is also shown in Fig.~4.
While a complete radio ``light curve'' is not yet available for \chicyg, 
there are two measures of its 3.6~cm wavelength emission made with the
NRAO\footnote{NRAO is a facility of the National Science Foundation operated under cooperative agreement by Associated
Universities, Inc.} Very Large Array: 
$0.27\pm0.04$~mJy at $\phi=0.21$ (Reid \& Menten) and $0.19\pm0.04$~mJy 
at $\phi=0.43$ \citep{rmg01}.  We converted these flux densities to 
magnitude units and plot them in Fig.~4.  
The model, adjusted only to optical and infrared data,
predicts the strength of the radio emission reasonably well. 
Also, the model predicts only a small variation of the radio flux density 
over the stellar cycle, which is consistent with that observed for 
other Mira variables (\oceti, R Leo, and W Hya) where better phase coverage
exists.

     We have shown that a simple model can successfully
fit the optical, infrared, and radio light curves of \chicyg.  
However, the temperature and density profiles of Mira variables 
are likely far more complex.  
So, why is our simple model so successful?  
One might be concerned that the model fits the data well only
because it has many adjustable parameters.  
Indeed, the model has 9 parameters.  
However, all of the parameters are 
physical ones, required by the problem.
The stellar distance comes from \citet{fea89} and the stellar mass
was arbitrarily set at 1~\Msun; neither were adjusted.  The two
stellar temperature and radius parameters are constrained mostly by the
infrared light curves.  The visual extinction only shifts the 
entire light curve up or down; the value adopted is small and not 
important.  This leaves only two parameters that can be used to 
fit the visual light curve: the oxygen depletion
and a single atmospheric density parameter ($n_*$).
So, the good fit of the model visual light curve to the observed data 
cannot be a result of over-parameterization. 

     How sensitive is the model visual light curve to the precise
nature of the temperature and density profiles?
In order to investigate this question,
we changed the density profile from an exponential form, appropriate
for a hydrostatic atmosphere, to a power-law form, appropriate for
a smoothly expanding atmosphere.   Using temperature and density profiles 
proportional to $r^{-1/2}$ and $r^{-5}$, respectively (see Fig.~3),
we were able to reproduce the large amplitude of the visual light curve 
of \chicyg\ nearly as well as the exponential model, while also obtaining
satisfactory agreement with the infrared and radio data.  
This suggests that the model light curves are somewhat robust to 
changes in the forms of the temperature and density profiles.  
Details of the variation of temperature and density 
with radius do not strongly affect the amplitude of the visual light curve, 
provided that sufficient opacity occurs near minimum light at radii where the 
temperature falls to a low value as discussed in \S2.

\section{Comments and Caveats}

     In some previous attempts to explain the phenomenon of Mira variables, 
the effects of TiO absorption have been considered as ``extinction''---that 
is simply absorbing the stellar light.  
However, this explanation is incomplete,
because the medium containing the absorbing material also re-emits light.
Consider, for example, the Sun.  The solar photosphere has a strong opacity 
source from the reaction ${\rm H}^- \rightarrow {\rm H} + {\rm e}^-$ 
\citep{wil39}; indeed, essentially all visible Sun light comes 
from the formation of H$^-$.  However, since density falls steeply with
radius in the Sun, the region of high opacity is confined to a very 
small range of radii.  Across this region, the 
temperature changes only slightly, and the visible light from the Sun
departs only marginally from a blackbody at a single temperature.  
Thus, strong absorption of light in a stellar photosphere is not alone 
sufficient to change the visual intensity markedly.  
The absorption must happen over a large range in radius, 
where the temperature changes significantly, if it is
to dramatically change the visual intensity.  This occurs in Mira variables, 
owing to molecular (e.g, TiO) absorption, 
but it does {\bf not} occur in the Sun, even though
there is equally strong H$^-$ absorption. 

     Our simple model cannot explain all details of the wide 
variety of light curves observed.  Some Miras exhibit asymmetric light curves, 
which decay slowly and rise rapidly toward maximum light; others display 
secondary ``humps" in their curves.  Also, Mira variables are not 
perfectly spherical---some are elongated by 10\% or more \citep{kar91}.  
Photospheric shocks and mass outflow generally tend to soften the 
density profile (i.e., increase the density at larger radii) above 
that obtained in pure hydrostatic equilibrium.   On the other hand, 
there is clear evidence that Mira atmospheres are not smoothly expanding.
Regions of large-scale infall are indicated by red-shifted absorption lines 
\citep{reid76,rd76,fg84} and they have been directly observed in the 
proper motions of some SiO masers \citep{dk99}.  
Thus, the combination of outflowing and infalling regions, 
perhaps better described as convective motions, 
when averaged over the entire atmosphere, might result in an average 
density profile that deviates less from a hydrostatic profile than 
expected from pure outflow.  This needs to be studied.
Still, explaining many details of Mira emissions requires more
sophisticated modeling.  
The simple hydrostatic model used here was chosen for computational ease, 
since this paper is concerned with explaining the basic physical reasons 
for the large amplitude of the visual light curve and not with constructing 
a detailed atmospheric model.  Ultimately, full dynamical models should be 
applied to this problem; these models should be constrained by observed light 
curves, in addition to observed spectra and measured sizes.

     Our simplifying assumption, that the visual opacity is dominated by TiO, 
is only a first approximation.  Other sources contribute to the overall 
opacity.  For example, stars of spectral class S exhibit strong bands from ZrO.
Indeed, \chicyg\ has been variously classified as an M-type (strong TiO 
bands) and an S-type (strong ZrO bands in addition to TiO) 
by different observers at different epochs. 
Including sources of opacity beyond TiO would decrease somewhat the required 
total density to achieve similar total opacity.  
This would further enhance the 
effects leading to the large amplitude of the visual light curves and make 
our explanation for these light curves more robust.   Some carbon-rich Mira 
variables of spectral class N and R are also known to have 
large amplitude light curves \citep{mer40}.  For these stars, essentially
all oxygen is in CO and metallic oxides, such as TiO, cannot form abundantly. 
Thus, the explanation for the behavior of the light curves of carbon-rich 
Mira variables probably involves strong opacity from carbon-bearing molecules 
such as C$_2$ and CN \citep{rg02}.    
Finally, dust can form at temperatures below $\sim1500 - 1200$~K in
carbon or oxygen rich Miras.  Indeed, dust shells are directly observed in 
some Miras, usually at radii greater than about $3R_*$ \citep{dan94}, 
and these can also contribute to the extinction of optical light 
(e.g., Winters \etal\ 1994).

     In conclusion, we mention that strong evidence for our basic picture 
comes from interferometric observations at visible and radio wavelengths.
Interferometric observations in the visible show large changes 
in the apparent size of the star for wavelengths in and out of
TiO absorption bands \citep{lab77,thb99,young00}.  
For example, imaging observations made with the Palomar 5-m telescope 
clearly show a ``shell" of high opacity at $\approx2R_*$ in a strong TiO 
absorption band \citep{han92}.
Independent observations at radio wavelengths \citep{rm97} directly measure the 
(electron) temperature at $\approx2R_*$ to be $\approx1500$~K.  This is
within the range needed to cause the large changes in visual light as
outlined in this paper.  Thus, based on direct observational evidence, 
the simplicity of the mechanism described above, and the ability of our models
to reproduce robustly the essential characteristics of Mira light curves, 
we feel that the physical picture outlined here may resolve the long-standing 
problem of the large amplitude of the light curves of Mira variables.

\acknowledgments
We thank R. Kurucz for providing the TiO cross-section data and
the AAVSO for the visual data on $\chi$~Cygni from their International 
Database, which comprises observations by variable star observers worldwide.

\clearpage

\begin{figure}
\epsscale{0.9}
\plotone{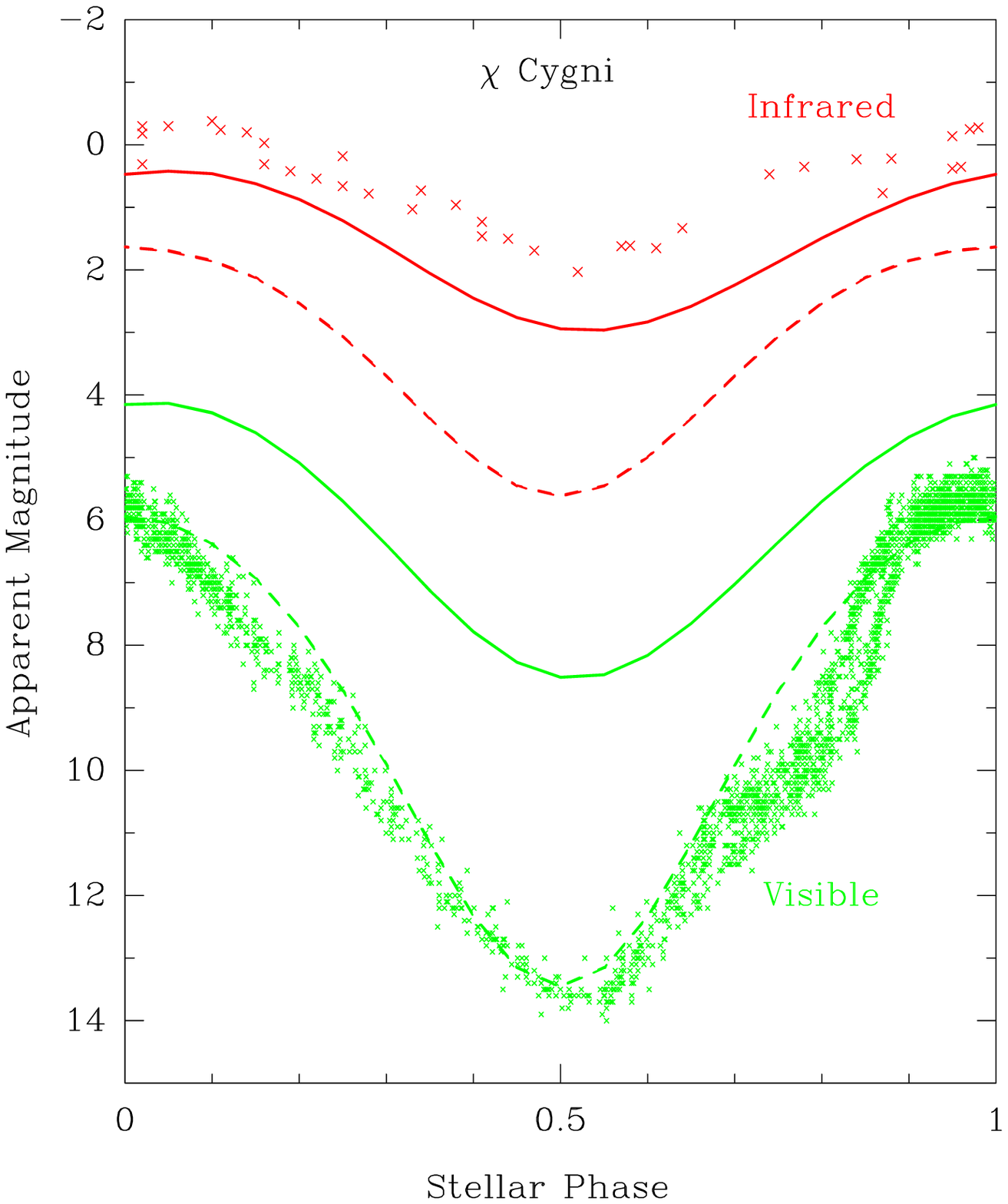}
\caption{
Observed infrared ({\it red crosses}), and visible ({\it green crosses})
light curves for the Mira variable $\chi$~Cygni.  (See text for references.) 
Solid lines are for a black--body model in which the temperature and
radius of the star vary as determined by \citet{pn33}.  Dashed lines
are for a black--body with a very large change in temperature: from
2000~K at maximum to 1240~K at minimum light.  Note that black--body models
cannot simultaneously fit the visible and infrared light curves.
\label{fig1}}
\end{figure}

\clearpage

\begin{figure}
\epsscale{1.0}
\plotone{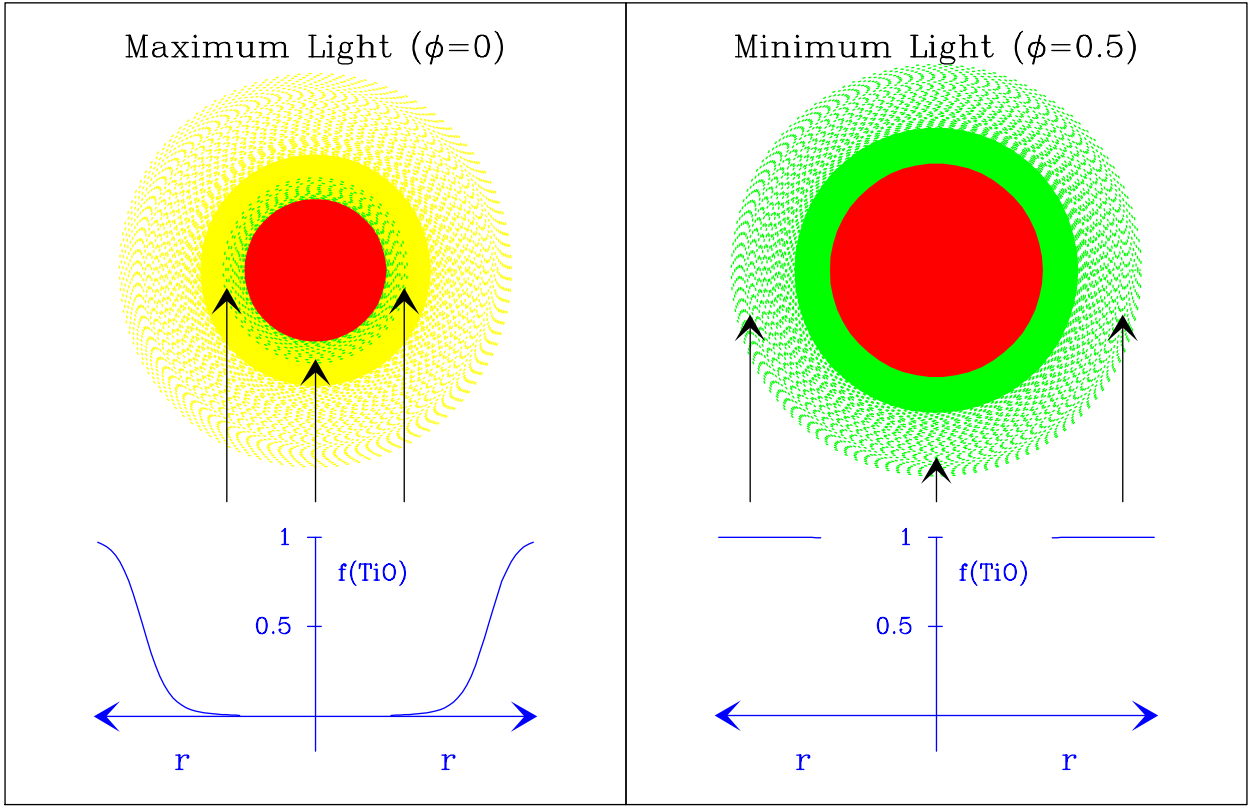}
\caption{
A schematic depiction of the change in visual appearance of a Mira 
variable star at maximum ({\it left-hand panel}) and minimum 
({\it right-hand panel}) light.   The star, shown in {\it red,} is
smaller and hotter at maximum light than at minimum light.
At maximum light, the extended atmosphere of the star 
(shown as {\it yellow}) is partially transparent at visual wavelengths, 
and one sees almost down to the stellar surface (indicated with {\it arrows}).
Near minimum light, the temperature of the star has declined and 
metallic oxides, such as TiO (shown as {\it green}), form throughout the 
extended atmosphere.  The fraction of Ti in TiO, f(TiO), as a function of 
radius is plotted in {\it blue}.  
Near minimum light, TiO forms with sufficient 
density at a radius of $\approx1.8R_*$ to become opaque to visible light.  
At this radius, the temperature can be very low, and almost all radiation 
is in the infrared.  Since little visible light emerges, the star can almost 
disappear to the human eye.  
\label{fig2}}
\end{figure}

\clearpage 

\begin{figure}
\epsscale{0.80}
\plotone{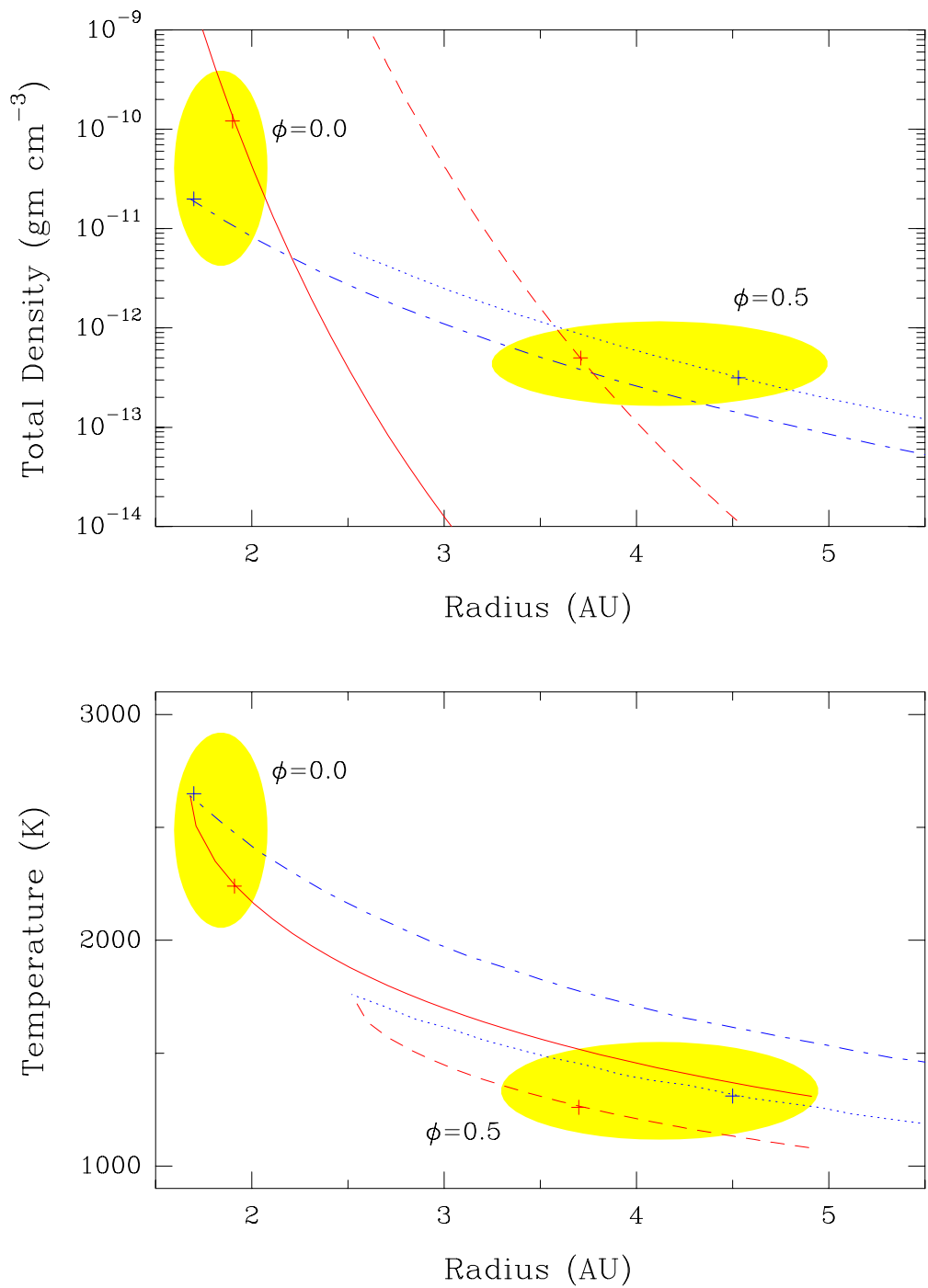}
\caption{
Models of density ({\it upper panel}) and temperature ({\it lower panel}) 
versus radius in the extended stellar photosphere of \chicyg.  
The {\it red} solid and dashed lines refer to a hydrostatic model 
at maximum and minimum visible light.  
The {\it blue} dash-dotted and dotted lines refer to an alternative, 
power-law, model at maximum and minimum visible light.  
The {\it yellow} ellipses, labeled with
visual phase ($\phi$), indicate roughly the regions of parameter
space that are likely to yield visual and infrared magnitudes similar to
those observed.  Pluses ($+$) indicate radii at which the visual optical depth
reaches unity in the models.   
\label{fig3}}
\end{figure}

\clearpage 

\begin{figure}
\epsscale{0.9}
\plotone{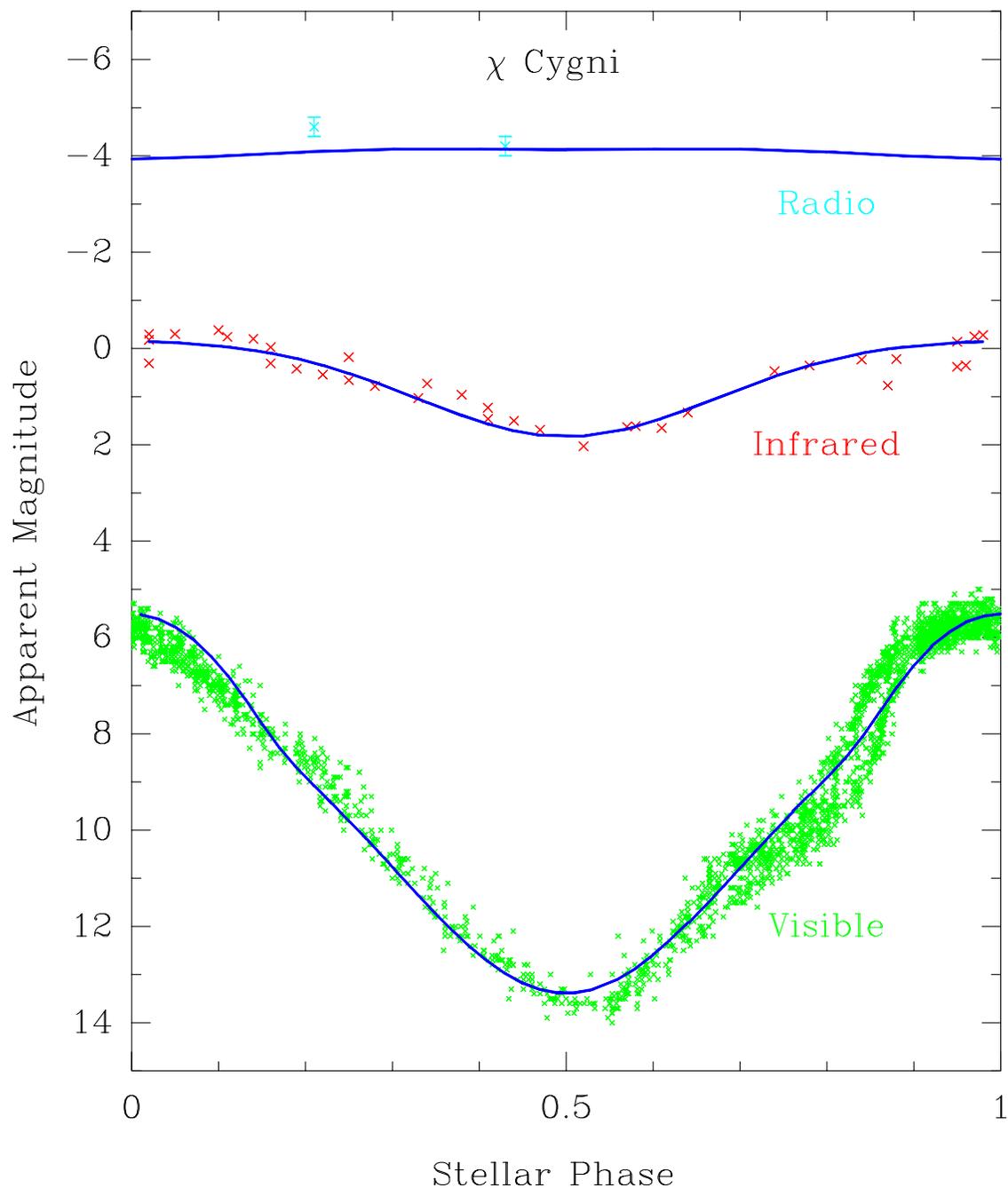}
\caption{
Observed radio ({\it cyan}), infrared ({\it red}), and visible ({\it green})
light curves for the Mira variable $\chi$~Cygni.  (See text for references.) 
Solid {\it blue} lines are from a simple model of an
oxygen--rich Mira variable with an extended atmosphere. 
Near stellar phase of 0.5, molecular absorption, mostly by TiO, 
enlarges the apparent visible size of the star to nearly $2R_*$.  
The low atmospheric temperatures ($\approx1400$~K) at this radius 
result in an $\approx8$ magnitude decrease in visible light compared 
to maximum light.  
The infrared data are insensitive to molecular absorption and respond mostly 
to the central pulsating star.  The radio emission is sensitive to the 
free electron density, and high radio opacity, owing to H$^-$ free--free 
interactions, occurs near $\approx2R_*$ throughout the stellar cycle. 
\label{fig4}}
\end{figure}

\end{document}